\documentclass[sigconf]{acmart}

\usepackage[utf8]{inputenc}
\usepackage{booktabs}
\usepackage{adjustbox}
\usepackage{graphicx}
\usepackage{subcaption}
\usepackage{multirow}
\usepackage{enumerate}
\usepackage{enumitem}
\usepackage{bbm}
\usepackage{amsmath}
\usepackage{graphicx}
\usepackage{amsfonts}
\usepackage{amssymb}
\usepackage{url}
\usepackage{acronym}
\usepackage{color}
\usepackage{mathtools}
\usepackage{tkz-graph}
\usepackage{tikz}
\usetikzlibrary{backgrounds}
\usetikzlibrary{trees,positioning,arrows}
\usepackage{wrapfig}
\usepackage[ruled,linesnumbered,lined,boxed]{algorithm2e}
\usepackage{diagbox}
\usepackage{array}

\newtheoremstyle{mystyle}
  {}
  {}
  {}
  {}
  {\scshape}
  {.}
  { }
  {}

\theoremstyle{mystyle}

\newcommand{\RQ}[2]{
    \begin{description}[topsep=0pt,nosep=0pt, leftmargin=0.75cm, noitemsep,nolistsep]
    \phantomsection\label{section:setup:rq#1}
    \item[RQ#1] #2
    \end{description}
}

\newcommand{\RQRef}[1]{\textbf{\hyperref[section:setup:rq#1]{RQ#1}}}

\let\cal\mathcal

\title{Technology Assisted Reviews: Finding the Last Few Relevant Documents by Asking Yes/No Questions to Reviewers}

\author{Jie Zou}
\affiliation{%
  \institution{University of Amsterdam}
  \city{Amsterdam}
  \country{The Netherlands}}
\email{j.zou@uva.nl}

\author{Dan Li}
\affiliation{%
  \institution{University of Amsterdam}
  \city{Amsterdam}
  \country{The Netherlands}}
\email{d.li@uva.nl}

  \author{Evangelos Kanoulas}
\affiliation{%
  \institution{University of Amsterdam}
  \city{Amsterdam}
  \country{The Netherlands}}
\email{e.kanoulas@uva.nl}

\renewcommand{\shortauthors}{J. Zou et al.}
\renewcommand{\shorttitle}{}

\begin{document}

\begin{abstract}
The goal of a technology-assisted review is to achieve high recall with low human effort. Continuous active learning algorithms have demonstrated good performance in locating the majority of relevant documents in a collection, however their performance is reaching a plateau when 80\%-90\% of them has been found. Finding the last few relevant documents typically requires exhaustively reviewing the collection. In this paper, we propose a novel method to identify these last few, but significant, documents efficiently. Our method makes the hypothesis that entities carry vital information in documents, and that reviewers can answer questions about the presence or absence of an entity in the missing relevance documents. Based on this we devise a sequential Bayesian search method that selects the optimal sequence of questions to ask. The experimental results show that our proposed method can greatly improve performance requiring less reviewing effort.
\end{abstract}



\copyrightyear{2018} 
\acmYear{2018} 
\setcopyright{acmlicensed}
\acmConference[SIGIR'18]{41st International ACM SIGIR Conference on Research and Development in Information Retrieval}{July 8--12, 2018}{Ann Arbor, MI, USA}
\acmBooktitle{SIGIR '18: 41st International ACM SIGIR Conference on Research and Development in Information Retrieval, July 8-12, 2018, Ann Arbor, MI, USA}
\acmPrice{15.00}
\acmDOI{10.1145/3209978.3210102}
\acmISBN{978-1-4503-5657-2/18/07}

\maketitle

\section{Introduction}

A Technology-Assisted Review (TAR) aims at locating all relevant documents in a collection (``total recall'') while minimizing manual reviewing effort. TAR has been successfully applied in a variety of high-recall tasks such as conducting systematic reviews in evidence-based medicine~\cite{OMara:2015}, electronic discovery in the legal proceedings~\cite{Cormack:2014}, creating test collections for Information Retrieval (IR) evaluation~\cite{Sanderson:2004}. 

O'Mara-Eves~\cite{OMara:2015} provides a detailed survey of machine learning methods used in TAR. Active learning techniques, which iteratively improve the accuracy of the predictions through interaction with reviewers, achieve state-of-the-art performance. In particular, Cormack and Grossman~\cite{Cormack:2014, Cormack:2017} have proposed the Baseline Model Implementation (BMI), a continuous active learning (CAL) algorithm, which has been evaluated in a number of high-recall tasks as the best performing algorithm~\cite{Grossman:2016, Kanoulas:2017}. BMI identifies an initial set of documents to be reviewed by experts to be used as an initial training set for learning a logistic regression model. The logistic regression algorithm predicts the relevance of the remaining of the documents. A set of top-scored documents is returned to assessors for labeling. The labeled documents are added back to the initial training set and the model is being retrained. 
While CAL algorithms have demonstrated their ability to efficiently find relevant documents in a collection~\cite{Cormack:2014, Grossman:2017}, recall typically reaches a plateau of 80\%-90\% after reviewing and labeling 30\%-40\% of the collection~\cite{Kanoulas:2017}. 
Finding the last few relevant documents requires reviewing almost the entire collection.


The goal of this work is to efficiently retrieve these last few relevant documents. 
Our hypothesis is that asking direct questions to reviewers will allow an algorithm to discover the missing documents faster than when requesting relevance feedback on documents through continuous active learning. Hence, we propose a Sequential Bayesian Search~\cite{Wen:2013} based method (SBSTAR), which locates the missing relevant documents efficiently by directly querying reviewers about significant pieces of information expected to appear, or not, in the relevant documents.
Our framework applies CAL up to a certain level of effort, in terms of documents reviewed. Then it switches to SBSTAR to directly ask questions to reviewers.
SBSTAR first identifies a pool of questions to be asked. In this work we focus on questions about the expected presence of an entity in the missing relevant documents. Hence, entities found in the corpus constitute the pool of available questions. SBSTAR then constructs a prior belief over document relevance on the basis of the ranking model trained by CAL. Then, it applies Generalized Binary Search (GBS) over entities to find the entity that dichotomizes the probability mass of document relevance. After each question is being answered by the reviewer a posterior belief is obtained to be used for the selection of the next question.

The main contribution of this paper is two-fold: (1) A method to construct a set of questions to be asked to the reviewers in terms of entities contained in the documents of the collection; (2) A novel interactive method, which directly queries reviewers about the expected presence of an entity in relevant documents, and updates the prior belief on document relevance at every round of interaction. 
To the best of our knowledge this is the first work that attempts to ask explicit questions to reviewers for the purpose of achieving total recall that goes beyond document relevance feedback. The evaluation results show that our approach can significantly reduce human effort, while achieve high recall.


\section{Methodology}
\label{sec:meth}

In this section, we provide a detailed description of the proposed method, which consists of two parts: (a) the construction of a pool of questions, and (b) the SBSTAR method to sequentially select questions to be asked to a reviewer towards finding the missing relevant documents.

\subsection{Question Pool Construction}
We consider entities to be the most vital source of information in text, an assumption made in previous work~\cite{rosen2004author,erosheva2004mixed}. Based on this assumption we focus on generating questions about the presence or absence of an entity in the relevant documents. We use TAGME~\cite{Ferragina:2010} to annotate entities in documents, and represent documents by a vector of entities. The algorithm asks a sequence of questions of the form “Are the documents you are interested in about [entity]?” to locate the target document of interest. We allow reviewers to respond with “yes”, “no”, and “not sure”, with the latter ensuring that reviewers are not forced to make erroneous choices when they are not certain about their answer.

\subsection{Sequential Bayesian Search for TAR}

The SBSTAR algorithm\footnote{https://github.com/jiezou0806/SBSTAR} is provided in Algorithm~\ref{algo}. The input to our algorithm is the document collection, ${\cal D}$, the set of annotated entities in the documents, ${\cal E}$, a prior belief, $\mathbb{P}_0$, which we model as a Dirichlet distribution parametrized by $\alpha$, and the number of questions to be asked, $N_q$. The initial $\alpha$ of the prior belief $\mathbb{P}_0$ is calculated by using the probability of a document being relevant provided by the CAL trained logistic regression.
%
We assume that there is a set of target relevant documents $d^* \in \mathcal{D}$. The reviewer preferences for the documents are modeled by a probability distribution $\pi^*$ over documents $\mathcal{D}$, and the target documents are drawn i.i.d. from this distribution. We also assume that there is a prior belief $\mathbb{P}_0$ over the reviewer preferences $\pi^*$, which is a probability density function over all the possible realizations of $\pi^*$. The system updates its belief when an reviewer's answer to a question is observed, which is sampled i.i.d. from $\pi^*$. First, we compute the certainty-equivalent reviewer preference $\pi^*_l(d)$. Let $\mathbb{P}_l$ be the system’s belief over $\pi^*$ in the $l$-th question, then
\begin{equation}
\pi^*_l(d) = \mathbb{E}_{\pi \sim \mathbb{P}_l} [\pi(d)] \quad \forall d \in \mathcal{D}
\end{equation}

After that, we use GBS to find the entity, $e_l$, that best splits the probability mass of the predicted document relevance, we ask whether the entity $e_l$ is present in the target document set, $d^*$, observe the reply $e_l(d^*)$, and remove $e_l$ from the entity pool. Then we update the system's belief $\mathbb{P}_l$ using Bayes' rule. Since the certainty-equivalent reviewer preference $\pi^*$ is a multinomial distribution over documents $\mathcal{D}$, we model the prior, 
$\mathbb{P}_0$, by the conjugate prior of the multinomial distribution, i.e., the Dirichlet distribution, with parameter $\alpha$. Further, we define the indicator vector $Z_l(d) = \mathbbm{1}\{e_l(d) = e_l(d^*)\}$, where $d^*$ represents the target documents. From Bayes' rule, the posterior belief at the beginning of question $l$ is: 
\begin{equation}
\mathbb{P}_l = Dir(\alpha + \sum_{j = 0}^{l-1} Z_j)
\end{equation}

From the properties of the Dirichlet distribution, then we have:
\begin{equation}
\pi^*_l(d) = \mathbb{E}_{\pi \sim \mathbb{P}_l} [\pi(d)] = \frac{\alpha(d)+ \sum_{j=0}^{l-1} Z_j(d)}{\sum_{d' \in \mathcal{D}} (\alpha(d')+ \sum_{j=0}^{l-1} Z_j(d'))}
\end{equation}

where $\alpha(d)$ is the i-th entry of $\alpha$, which corresponds to document d. Therefore the certainty-equivalent reviewer preference $\pi^*_l$ can be updated by counting and re-normalization. After the last question is being asked, the relevance ranking list is generated based on the reviewer preference $\pi^*_{N_q}$ over the remaining documents.

\begin{algorithm}[t]
\SetKwInOut{Input}{input}
\Input{A document set, ${\cal D}$, the set of annotated entities in the documents, ${\cal E}$ , a prior belief over document relevance, $\mathbb{P}_0$, and a number of questions to be asked, $N_q$}
\BlankLine
\ForEach{topic}{
  $l \leftarrow 1$
  
  \While{$l \le N_q$}{
    Compute the certainty-equivalent reviewer preference: $\pi^*_l(d) = \mathbb{E}_{\pi \sim \mathbb{P}_l}[\pi(d)] ~ \forall d \in {\cal D}$ \ 
    
    Using GBS to find the optimal target entity:\    
  
    $e_l = \arg\min_{e}|\sum_{d \in {\cal D}} (2 \mathbbm{1}\{e(d)=1\} - 1)\pi^*(d)|$\
    
    Ask the question about $e_l$ and observe the reply $e_l(d^*)$\    

    Remove $e_l$ from entity pool\    

    $l \leftarrow l + 1$\

    Update the system's belief $\mathbb{P}_l$ using Bayes' rule:\
    $\mathbb{P}_{l+1}(\pi) \propto \pi(d)\mathbb{P}_l(\pi) ~ \forall \pi$\
  }
}
\caption{SBSTAR}
\label{algo}
\end{algorithm}

\section{Experiments and Analysis}
Through the experiments conducted in this work we aim to answer the following research questions:
\RQ{1}{What is the impact of the CAL stopping point, after which SBSTAR is applied, as well as the impact of the number of questions asked?}
\RQ{2}{How effective is the proposed method in finding missing relevant documents compared to state-of-the-art algorithms?}

\subsection{Experimental setup}

\textit{Dataset.} In our experiments we use the collection released by CLEF 2017 e-Health Evaluation Lab~\cite{Kanoulas:2017}. The collection consists of $50$ topics and $266,967$ abstracts of MEDLINE articles, and the relevance judgments for each of these articles against the 50 topics.

\noindent\textit{Evaluation measures.} To quantify the quality of algorithms we use two of the official evaluation measures provided by CLEF 2017 e-Health Evaluation Lab~\cite{Kanoulas:2017}, Average Precision (AP) and last\_rel, that is the position of the last relevant document in the ranking, which to some extend quantifies the effort, in terms of reviewed documents, that is required to achieve total recall.

\noindent\textit{Simulating reviewers.} Our experimentation depends on reviewers responding to questions asked by our method. We simulate reviewers, that respond to the questions with full knowledge of whether an entity is present or not in the missing documents. Hence, we assume that a reviewer will respond with ``yes'' if an entity is contained in all missing relevant documents, ``no'' if an entity is absent from all missing relevant documents, and ``not sure'' for anything in between. We leave the development of noise-tolerant algorithms as future work.

\noindent\textit{Baselines.} We compare our method to three baselines, (1) \textbf{BMI}~\cite{Cormack:2017}, which is the state-of-the-art continuous active learning algorithm applied without any stopping criterion until the entire collection is reviewed, (2) \textbf{BMI + LR}, which applies BMI until a stopping point and then ranks the remaining of the collection on the basis of the trained logistic regression model, and (3) \textbf{BMI + Random}, which applies BMI until a stopping point and then randomly chooses the entities to ask questions about. When simulating reviewers we make a very strong assumption regarding the ability of a reviewer to precisely know whether an entity appears in all remaining relevant documents. While in the future we plan to relax this assumption, we still want to understand whether the proposed algorithm performs a sensible search over potential queries, hence the comparison with random selection method.

\subsection{The effect of the stopping point and the number of questions}

\begin{figure}
  \caption{Heatmap of the total effort required to reach 100\% recall. The total effort is naively defined as the sum of rank of the last relevant document and the number of queries asked. The total effort is shown as a function of the stopping point (stop ratio) and the number of questions asked. The more blue the heatmap the better the performance of the method. The boxes with ta white boundary box designate the optimal number of questions for the corresponding stopping point.}
  \includegraphics[width=1\columnwidth]{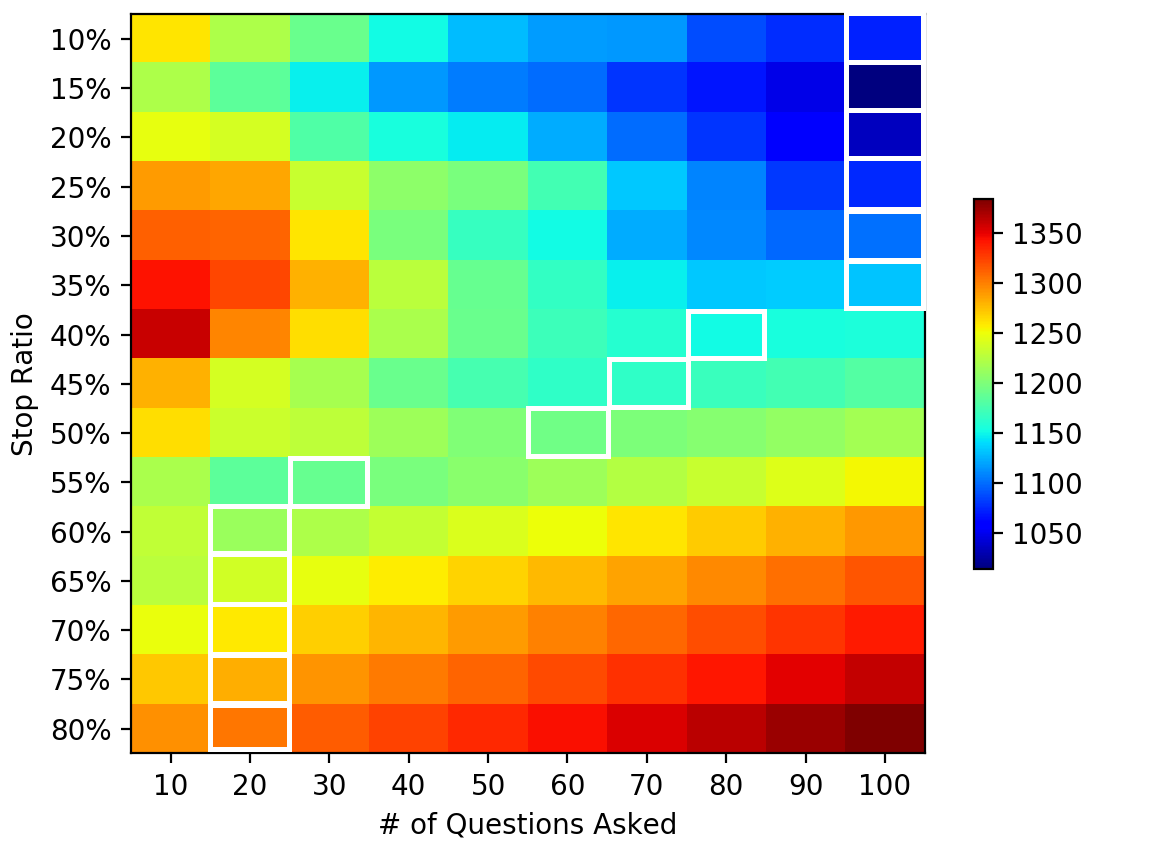}
  \label{fig:heatmap}
\end{figure}

In this section we answer \RQRef{1}. Our proposed method is parameterized by the stopping point of BMI and the number of questions to be asked to the reviewer. A number of approaches has been developed in identifying a good stopping point for continuous active learning algorithms, such as the ``knee'' method~\cite{Cormack:2017}, however we leave this as a free parameter in our experiments, to better understand its effect on the performance of our algorithm. We do the same for the number of question asked to the reviewer which range from 10 to 100. Figure~\ref{fig:heatmap} shows the heat map of the effort required to reach total recall. The x-axis is the number of questions asked and the y-axis the stopping point as a percentage of the collection shown to the reviewer by BMI. The effort is measured by two indicators: (a) the total number of documents that are required to be reviewed to reach total recall (i.e. the last\_rel measure); this includes both the documents ranked by BMI before the stopping point and the documents ranked by SBSTAR after the stopping point, and (b) as the number of questions asked by BMI. The effort is computed as the sum of the two numbers, by making the simplifying assumption that answering a question takes the same time as providing the relevance of a document. The optimal number of questions for each stopping point is indicated by the white boundary box.

As it can be observed the effort is increasing with the number of asked questions when the stop ratio is greater than or equal to 55\%, while the effort is decreasing when the stop ratio is less than or equal to 50\%. This is because there are very few missing relevant documents when the stop ratio is set to a high value, in which case asking many questions only leads to higher effort. Further, SBSTAR can effectively reduce the effort when the stop ratio is less than or equal to 50\%. The effort fluctuates over different stop ratio and the effort is relatively lower when stop ratio is between 15\% and 20\%, and between 45\% and 55\%. The lowest effort is achieved when stop ratio is 15\% and the number of asked questions is 100.

\subsection{The performance of the SBSTAR method}

\begin{table}[t]
\caption{Comparison of performance on MAP and last\_rel with different stopping points (in terms of the percentage of documents reviewed through BMI). For each stopping point the near-optimal number of questions were asked as indicated by the white-boundary boxes in Figure~\ref{fig:heatmap}.}
\label{table:2}
\centering
  \small
\begin{tabular}{p{0.08\columnwidth}<{\centering} | p{0.04\columnwidth}<{\centering} p{0.08\columnwidth}<{\centering}  p{0.08\columnwidth}<{\centering} p{0.1\columnwidth}<{\centering} |p{0.04\columnwidth}<{\centering} p{0.08\columnwidth}<{\centering}  p{0.08\columnwidth}<{\centering} p{0.1\columnwidth}<{\centering}  }
\hline
 & \multicolumn{4}{p{0.4\columnwidth}<{\centering}|}{MAP} & \multicolumn{4}{p{0.4\columnwidth}<{\centering}}{last\_rel}\\
\hline 
Stop Ratio & BMI & BMI+ LR &BMI+ Random & SBSTAR &BMI & BMI+ LR & BMI+ Random & SBSTAR\\
\hline 
10\% & 0.167 & 0.164 & 0.235 & \textbf{0.606} & 993 & 1400 & 1307 & \textbf{811} \\
15\% & 0.105 & 0.134 & 0.225 & \textbf{0.687} & 929 & 1270 & 1153 & \textbf{617} \\
20\% & 0.104 & 0.124 & 0.257 & \textbf{0.740} & 820 & 1084 & 990 & \textbf{491} \\
25\% & 0.063 & 0.09 & 0.271 & \textbf{0.779} & 737 & 1038 & 949 & \textbf{412} \\
30\% & 0.049 & 0.11 & 0.278 & \textbf{0.769} & 749.3 & 1100 & 965 & \textbf{353} \\
35\% & 0.082 & 0.121 & 0.26 & \textbf{0.776} & 720 & 1114 & 968 & \textbf{279} \\
40\% & 0.053 & 0.083 & 0.282 & \textbf{0.68} & 814 & 1159 & 1027 & \textbf{255} \\
45\% & 0.036 & 0.069 & 0.326 & \textbf{0.683} & 785 & 947 & 872 & \textbf{181} \\
50\% & 0.043 & 0.094 & 0.369 & \textbf{0.806} & 644 & 690 & 651 & \textbf{154} \\
55\% & 0.139 & 0.117 & 0.244 & \textbf{0.831} & 545 & 605 & 589 & \textbf{58} \\
60\% & 0.1 & 0.035 & 0.093 & \textbf{0.897} & 760 & 925 & 892 & \textbf{34} \\
65\% & 0.004 & 0.003 & 0.004 & \textbf{1} & 1414 & 1742 & 1637 & \textbf{21} \\
70\% & 0.001 & 0.001 & 0.001 & \textbf{1} & 1079 & 1426 & 1383 & \textbf{21} \\
75\% & 0.002 & 0.001 & 0.004 & \textbf{1} & 734 & 1146 & 702 & \textbf{21} \\
80\% & 0.013 & 0.001 & 0.041 & \textbf{1} & 391 & 865 & 737 & \textbf{21} \\
\hline
Avg & 0.064 & 0.076 & 0.193 & \textbf{0.817} & 808 & 1101 & 988 & \textbf{249}\\
\hline
\end{tabular}
\end{table}

To answer \RQRef{2} we compare the effectiveness of our proposed method with the state-of-the-art baselines. Here, we calculate MAP and last\_rel only on the documents ranked after the stopping point, since we want to isolate the effectiveness of the proposed method. For each stopping point the optimal number of questions were asked by SBSTAR and Random, indicated by the white-boundary boxes in Figure~\ref{fig:heatmap}. The results of the comparison measured by MAP and last\_rel are shown in Table~\ref{table:2}. The best-performing values are shown in boldface. Our method outperforms BMI, BMI + LR, and BMI + Random both with respect to MAP and last\_rel. This clearly suggests that a theoretically optimal sequence of entity-centered questions can be rather effective. Table~\ref{table:1} provides an example of a sequence of questions session.

\begin{table}[t]
\caption{An example of a sequence of questions asked by SBSTAR.}
\label{table:1}
\begin{center}
\begin{tabular}{llr}
\hline
\multicolumn{3}{p{\columnwidth}}{Topic: Human papillomavirus testing versus repeat cytology for triage of minor cytological cervical lesions}\\
\hline
\multicolumn{3}{p{\columnwidth}}{Missing documents:} \\
\multicolumn{3}{p{\columnwidth}}{ID: 19116707, Title: Prevalence of human papillomavirus types 6, 11, 16 and 18 in young Austrian women - baseline data of a phase III vaccine trial.}\\
\multicolumn{3}{p{\columnwidth}}{ID: 19331088, Title: Cervical cytology screening and management of abnormal cytology in adolescents.}\\
\hline 
Question & Answer & Rank of Last Relevant  \\
Are the documents about ... &  & 988\\
Human Papillomavirus (HPV)&Yes & 430\\
women & Not Sure & 430\\
cervical cancer &Yes & 224\\
infection & Yes & 129\\
cancer& Yes & 44\\
development& No & 19\\
treatment& Not Sure & 19\\
disease& Yes & 6\\
clinic& No & 5\\
cervical& Yes & 2\\
\hline
\end{tabular}
\end{center}
\end{table}

\section{Conclusion and Future Work}
The focus of this work is achieving high recall in technology-assisted reviews. We propose a novel interactive method, SBSTAR, which directly queries reviewers on the presence or absence of an entity in missing relevant documents. Our framework applies continuous active learning on reviewers' relevance feedback until a certain percentage of documents has been reviewed and then switches to the proposed SBSTAR model to find the last few missing relevant documents. Experiments on the CLEF 2017 e-Health Lab demonstrate that the SBSTAR model can find the missing relevant documents efficiently, requiring minimal effort from reviewers.

In our work we make the assumption, that reviewers, when presented with an entity, they know, with 100\% confidence, whether the entity appears in all missing documents. This is a strong assumption. 
We leave the investigation of noise-tolerant algorithms, that will allow us to relax the assumption of 100\% confidence of reviewers when answering a query, as future work.
A second assumption made in this work is that answering a direct question about entities requires at most as much effort as judging the relevance of a document. To verify this assumption a user study is necessary, which we also leave as a future work.
The performance of the entity annotation algorithms affects the performance of our proposed method. In this paper, we use TAGME, however entity annotators that specialize to medical entities could yield improvements.
%

\bibliographystyle{ACM-Reference-Format}
\bibliography{bibfile} 

\end{document}